\documentclass[aps,prb,showpacs,twocolumn,preprintnumbers,amsmath,amssymb]{revtex4}
\begin{document}
\title{Comment on "Nature of the high - pressure tricritical point in MnSi"
}

\author{Sergei M. Stishov}

\email{sergei@hppi.troitsk.ru} \affiliation{Institute for High
Pressure Physics of Russian Academy of Sciences, Troitsk, 142190 Moscow
Region, Russia}

\date{\today}

\begin{abstract}
It is argued that M. Otero-Leal et al. [PRB 79, 060401 (2009)\cite{1}] wrongly identified the second order term of the Arrott equation with the coefficient at the quartic term of the Landau expansion, therefore  deriving absolutely unsupported conclusions on the phase diagram of MnSi.
\end{abstract}


\maketitle

Despite a long history of extensive studies of the magnetic phase transition in the helimagnet MnSi its phase diagram is not completely understood. Most important but still controversial issues concerning the phase diagram of MnSi are the existence, location and nature of a tricritical point on the phase transition line in MnSi, first suggested in Ref. \cite{2}. Since then until recently, the existence of the tricritical point has been never in doubt though its location was disputed in Ref.\cite{3}. Meanwhile a careful study of the magnetic phase transition in MnSi confirmed a first order nature of the transition at  least at ambient pressure \cite{4,5,6}, as was proposed long ago in Ref.\cite{7}. This observation reversed the situation. It was believed before that a second order phase transition in MnSi became first order at high pressure. Then the only conclusion could be made that the phase transition in MnSi turned to second order at high pressure. The latter of course is valid if a tricritical point does exist at the transition line. However new experiments at high pressure using helium as a pressure medium showed that the early claims on the existence of a tricritical point at the phase transition line, based on an analysis of behavior of magnetic susceptibility, are most probably a result of misinterpretation of the experimental data\cite{8}. As was said in Ref.\cite{8}:"The new measurements unambiguously show that the striking change of the magnetic susceptibility curve is a result of non-hydrostatic stresses in pressure media and has nothing to do with a change of character of the phase transition in MnSi". So the general conclusion was that the magnetic phase transition in MnSi continued to be first order in the whole pressure range studied \cite{8}.
In the paper \cite{1} an attempt was made to resolve the above mentioned issues based, as the authors said, on "a direct analysis of the magnetic phase transition under pressure". Under "a direct analysis" the authors implied  applications of the equation
\[H/M=a+bM^{2}+cM^{3}+...		 , (1)\]
where  $H$- magnetic field, $M$ -magnetization.
Equation (1), suggested by Arrott \cite{9}, is a typical mean field relation and can be readily derived from the Landau expansion \cite{10}. Normally the relation (1) is used to plot $H/M$ vs $M^{2}$  (Arrott plot), which would yield a straight line at the temperature of second order ferromagnetic phase transition when the critical fluctuations can be neglected. Eq. (1) needs to be modified to account for the contribution of critical fluctuations (see Ref.\cite{11}).
It has to be emphasized that a causal relationship  between the Landau expansion and Eq.1 limited to ferromagnetic materials, since only in this case the magnetization $M$  can serve as an order parameter with the magnetic field $H$  as a conjugate variable.
Neglecting this important circumstance the authors \cite{1} applied Eq.(1) to the helical magnet MnSi though it was known for years that the Arrott approach did not work in this case \cite{12}. The reason for that is quite obvious. An order parameter for a helical magnet is a varying spin density, which  is not conjugate to the magnetic field (indeed the magnetic field can not create a helical order). Details of the Landau approach to the phase transition in MnSi see for instance in Ref.\cite{7,13}.
Nevertheless the authors Ref.\cite{1} analyzed behavior of the coefficient $b$  in Eq.(1) as functions of magnetic field and pressure at temperatures slightly above the phase transition line. They again found  as in Ref.\cite{12} strongly nonlinear relationship between $H/M$  and $M^{2}$, but ignoring this specifics, they concentrated on variations of the coefficient $b$  in Eq.(1), which became negative at low magnetic fields at pressure above 3.5 kbar. Note that this fact would be relevant only in the ferromagnetic case. Then wrongly identifying  the coefficient $b$  with the coefficient at the quartic term of the Landau expansion they derived absolutely unsupported conclusions on the phase diagram of MnSi. Finally the authors' pretensions "to put an end to the controversy about the nature of the magnetic phase transition and its evolution with pressure in MnSi" do not look even slightly founded.

\end{document}